\documentclass{aa501}

\usepackage[tbtags,fleqn]{amsmath}
\usepackage{amsfonts}
\usepackage{graphicx}

\newcommand{\diff}{\mathrm d}

\newcommand{\mincir}{\raise
  -2.truept\hbox{\rlap{\hbox{$\sim$}}\raise5.truept \hbox{$<$}\ }}
\newcommand{\magcir}{\raise
  -2.truept\hbox{\rlap{\hbox{$\sim$}}\raise5.truept \hbox{$>$}\ }}

\begin{document}

\title{Boyle's law and gravitational instability}
\author{Marco Lombardi\inst{1} and Giuseppe Bertin\inst{2}}
\offprints{M. Lombardi}
\mail{lombardi@astro.uni-bonn.de}
\institute{%
  Instit\"ut f\"ur Astrophysik und Extraterrestrische Forschung,
  Universit\"at Bonn, Auf dem H\"ugel 71, D-53121 Bonn, Germany 
  \and
  Universit\`a degli Studi di Milano, Dipartimento di Fisica, via
  Celoria 16, I-20133 Milano, Italy}
\date{Received ***date***; accepted ***date***}

\abstract{%
  We have re-examined the classical problem of the macroscopic
  equation of state for a hydrostatic isothermal self-gravitating gas
  cloud bounded by an external medium at constant pressure.  We have
  obtained analytical conditions for its equilibrium and stability
  without imposing any specific shape and symmetry to the cloud
  density distribution.  The equilibrium condition can be stated in
  the form of an upper limit to the cloud mass; this is found to be
  inversely proportional to the power $3/2$ of a form factor $\mu$
  characterizing the shape of the cloud.  In this respect, the
  spherical solution, associated with the maximum value of the form
  factor, $\mu = 1$, turns out to correspond to the shape that is most
  difficult to realize.  Surprisingly, the condition that defines the
  onset of the Bonnor instability (or gravothermal catastrophe) can be
  cast in the form of an upper limit to the density contrast within
  the cloud that is independent of the cloud shape.  We have then
  carried out a similar analysis in the two-dimensional case of
  infinite cylinders, without assuming axisymmetry.  The results
  obtained in this paper generalize well-known results available for
  spherical or axisymmetric cylindrical isothermal clouds that have
  had wide astrophysical applications, especially in the study of the
  interstellar medium.  }

\maketitle

%

\section{Introduction}

Consider a static isothermal gas cloud embedded in a medium
characterized by an external pressure $p_\mathrm{b}$.  Under
equilibrium conditions, it is expected to satisfy the virial
theorem:
\begin{equation}
  \label{eq:1}
  2 K + W = 3 p_\mathrm{b} V \; ,
\end{equation}
where $K$ is the total internal kinetic energy associated with the
thermal motions of the gas molecules and $W$ is the total
gravitational energy.  From the relation $K = 3 M c_\mathrm{s}^2/2$,
we see that
\begin{equation}
  \label{eq:2}
  p_\mathrm{b} V = M c_\mathrm{s}^2 + \frac{1}{3} W < \frac{M k T}{m}
  \; .
\end{equation}
Here $c_\mathrm{s}$ is the thermal speed inside the cloud, given by
$c_\mathrm{s} = \sqrt{k T / m}$, where $k$ is the Boltzmann constant,
$T$ the cloud temperature, and $m$ the mean mass of the gas particles.
Hence, the presence of gravity softens the standard equation of state
$p_\mathrm{b} V = M k T / m$ for a gas cloud of volume $V$ (Boyle's
law) into one characterized by an \textit{effective temperature\/}
$T_\mathrm{eff} < T$.  Clearly, $T_\mathrm{eff}$ depends on the size,
the shape, and the temperature of the cloud.  In particular, for given
values of all other parameters, we could imagine density distributions
for which $W$ is a large negative quantity, so that $T_\mathrm{eff} <
0$; these distributions would not be compatible with the assumed
equilibrium conditions.  On the other hand, even when $T_\mathrm{eff}
> 0$, the equilibrium configuration might not be viable because it
could be unstable.

Within spherical symmetry, this line of thought has led to the
remarkable discovery (Bonnor 1956; see also Ebert 1955) that a
thermodynamic type of instability takes place when the density
contrast (between the center and the periphery of the cloud) exceeds a
certain threshold value.  This relatively simple conclusion has opened
the way to a variety of interesting ``applications", best exemplified
by the interpretation of the structure of weakly collisional stellar
systems (globular clusters) in terms of the onset of the so-called
{\it gravitational catastrophe} (see Lynden-Bell \& Wood 1968 and
following papers; in particular, see Spitzer 1987).  The underlying
reason for the instability can be traced to a basic property of
self-gravity, of being associated with a negative specific heat.  It
would be conceptually interesting to study whether and how the general
conclusions change for different cloud shapes.

A study of such well-posed mathematical problem appears to be timely.
Indeed, there has been enormous progress in the study of the cloudy
structure of the interstellar medium (e.g., Falgarone et al.\ 1992;
Falgarone et al.\ 1998; Lada et al.\ 1994; Alves et al.\ 1998).
Curiously, some of the properties that are often considered as
analytically useful in idealized theoretical models have been shown to
be, at least in some cases, also realistic.  In particular, while the
possibility of non-trivial shapes for gas clouds is clearly suggested
by observations of the cold interstellar medium (see, e.g., Loren
1989a, Lada et al.\ 1991), in some cases these observations
quantitatively support a simple interpretation in terms of pure
hydrostatic equilibrium (see, e.g., Alves et al.\ 2001).  Similarly,
it has long been recognized that cold clouds generally conform to the
virial equilibrium conditions (e.g., see Myers 1983; Loren 1989b;
Jijina et al.\ 1999).  Furthermore, although magnetic fields and
rotation are generally accepted to play a significant role in the
dynamics of gas clouds (see Goodman et al.\ 1995), there is evidence
that cloud shapes are decoupled from the presence of magnetic fields
(compare the case of the Taurus cloud, Heyer et al.\ 1987, with that
of the $\rho$ Ophiuchi cloud, Goodman et al.\ 1990) and that rotation
can often be insignificant (e.g., Walker et al.\ 1993); moreover,
statistical arguments applied to the distribution of measured axial
ratios for cloud cores suggest that those structures are intrinsically
prolate (David \& Verschueren 1987; Myers et al.\ 1991).

Therefore, we find it natural to address the issue of the stability of
an isothermal cloud in hydrostatic equilibrium, bounded by an external
medium characterized by a specified constant pressure $p_\mathrm{b}$,
in its most general case, that is beyond the well-known framework of
spherical symmetry.  Since we do not consider the role of magnetic
fields and, especially, non-thermal motions, we limit our discussion
to the scale of cloud cores, typically a few tenths of a parsec.
Still, a study on such scales is very important if we wish to
understand the process of star formation.  In contrast with previous
analytical investigations, we then let the shape and density
distribution of the cloud deviate \textit{significantly\/} from the
spherical shape.  The origin of these shapes should be traced to the
presence of external forces (tidal fields and/or other boundary
forces).  Here we will not be concerned about how a specific shape can
be realized, appealing to the empirical fact that observations
demonstrate that cloud cores find the way to settle in a variety of
shapes.  Note that filamentary structures are very common (e.g., Loren
1989a; Bally et al.\ 1987), and often each filament contains several
distinct subcondensations, sometimes periodically spaced (Dutrey et
al.\ 1991).

By generalizing the study by Bonnor (1956), in this paper we obtain an
expression for the compressibility $\partial p_\mathrm{b} / \partial
V$ at constant temperature and fixed shape and thus reconstruct the
isothermal curves in the $(V, p_\mathrm{b})$ plane.  These are
parameterized by a \textit{dimensionless form factor} $\mu$, a
quantity that depends only on the shape of the cloud, but otherwise
follow the qualitative behavior characteristic of the self-gravitating
isothermal sphere; in particular, no equilibrium is available above a
certain value of the boundary pressure.  Our analysis quantifies how
such equilibrium condition depends on $\mu$ and shows that the sphere,
being associated with the maximum value for the form factor ($\mu =
1$), corresponds to the shape for which equilibrium is most difficult
to realize.  We then proceed to address the condition for the Bonnor
instability and show that it can be cast in a form that is actually
independent of the cloud shape.

The paper is organized as follows.  The relevant basic relations are
introduced in Sect.~\ref{sec:basic-relations}.  In
Sect.~\ref{sec:boyles-law} we derive the equivalent of Boyle's law for
self-gravitating clouds.  The implications of the results obtained for
the equilibrium and stability of the cloud are clarified in
Sect.~\ref{sec:bonnor-instability} and are summarized in two simple
conditions (Eqs.~\eqref{eq:34} and \eqref{eq:35}).  In
Sect.~\ref{sec:boyles-law-cylinders} we present analogous results for
(infinite) cylinders.  Finally, in Sect.~\ref{sec:conclusions}, we
briefly summarize the main results obtained in this paper.

\section{Basic relations}
\label{sec:basic-relations}

We summarize here the basic relations that will be used through this
paper.

Let us call $\rho(\vec x)$ the mass density distribution and $p(\vec
x)$ the (scalar) pressure of a cloud at point $\vec x$.  If the cloud
is in hydrostatic equilibrium, then
\begin{equation}
  \label{eq:3}
  \nabla p = -\rho \nabla \Phi \; ,
\end{equation}
where $\Phi(\vec x)$ is the Newtonian gravitational potential,
described, inside the cloud, by the Poisson equation
\begin{equation}
  \label{eq:4}
  \nabla^2 \Phi = 4 \pi G \rho \; .
\end{equation}
Furthermore, if the cloud is isothermal we can write
\begin{equation}
  \label{eq:5}
  p = c_\mathrm{s}^2 \rho \; .
\end{equation}
Taking $T = 10 \mbox{ K}$ and a cloud made of molecular hydrogen, we
find a typical value of $c_\mathrm{s} \simeq 200 \mbox{ m s}^{-1}$.

Equations (\ref{eq:3}--\ref{eq:5}) combined together lead to
Emden's equation (Emden 1907):
\begin{equation}
  \label{eq:6}
  \nabla \cdot \left( \frac{\nabla \rho}{\rho} \right) = 
  \nabla^2 \ln (\rho / \rho_0) = - \frac{4 \pi G}{c_s^2} \rho \; ,
\end{equation}
where $\rho_0$ is a characteristic density (e.g., the maximum density
of the cloud).  A useful property of Eq.~\eqref{eq:6} that is the
basis of the following analysis is the \textit{homology theorem\/}
(Chandrasekhar 1967): If $\rho(\vec x)$ is a solution of this
equation, then $\lambda^2 \rho(\lambda \vec x)$ is also a solution.
In fact, we can write Eq.~\eqref{eq:6} in dimensionless form by
introducing the variables
\begin{equation}
  \label{eq:7}
  u = \frac{\rho}{\rho_0}
\end{equation}
and $\vec\xi = s \vec x$, where $s^{-1}$ is a suitable ``Jeans
length'' (cf.\ Jeans 1929)
\begin{equation}
  \label{eq:8}
  s = \sqrt{\frac{4 \pi G \rho_0}{c_\mathrm{s}^2}} \; .
\end{equation}
Hence $u$ is solution of the equation
\begin{equation}
  \label{eq:9}
  \nabla^2_\xi \ln u(\vec \xi) = -u(\vec \xi) \; ,
\end{equation}
where $\nabla^2_\xi$ is the Laplacian with respect to the variable
$\vec \xi$.  In the following we will reserve the subscript $b$ for
quantities measured at the boundary of the cloud.

\section{Boyle's law}
\label{sec:boyles-law}

In this section we will derive Boyle's law for a self-gravitating
cloud of arbitrary shape subject to a uniform external pressure
$p_\mathrm{b}$.  In particular, we will obtain an expression for the
derivative $\partial p_\mathrm{b} / \partial V$ at constant
temperature and fixed shape, for a cloud of given mass.  A similar
calculation had been performed by Bonnor (1956; see also Ebert 1955)
in the case of a spherical cloud.  Here we imagine that the cloud is
embedded in an external medium in an arbitrary dynamical state.  Our
main assumption is that such an arbitrary state enforces a non-trivial
geometry for the cloud, while the cloud has had time to settle into a
hydrostatic equilibrium and is thus described by Eq.~\eqref{eq:9}.
The possibility of such non-trivial shapes is clearly suggested by
observations of the cold interstellar medium (see, e.g., Loren 1989a,
Lada et al.\ 1991), which in some cases also support the hydrostatic
hypothesis (see, e.g., Alves et al.\ 2001).
 
Calculations will be carried out using the following scheme: We will
first calculate the change of the cloud volume corresponding to a
change of the density parameter $\rho_0$; then we will relate the
variation of the density parameter $\rho_0$ to the variation of the
pressure $p_\mathrm{b}$ at the boundary of the cloud.  The shape of
the cloud is taken to remain unchanged.

Since the cloud is in equilibrium with an external pressure
$p_\mathrm{b}$, the boundary of the cloud is defined by an iso-density
surface characterized by $\rho_\mathrm{b} = p_\mathrm{b} /
c_\mathrm{s}^2$.  When the external pressure changes, the cloud
boundary remains an iso-density surface.  In other words, the
transformations considered involve only a change of $\rho_0$ (and thus
of $s$) in $\rho(\vec x)$, while $u(\vec \xi)$ remains unchanged.
Finally, we note that, since the boundary is an iso-density surface,
$\nabla \rho$ and $\nabla_\xi u$ are normal to the cloud boundary.

\subsection{Perturbations at constant mass}
\label{sec:constance-mass}

Suppose that the density parameter $\rho_0$ changes by a quantity
$\delta \rho_0$.  Then the cloud density distribution will change by
\begin{equation}
  \label{eq:10}
  \delta\rho(\vec x) = \delta \rho_0 u(s \vec x) + \delta \rho_0
  \frac{\diff s}{\diff \rho_0} \rho_0 \nabla_\xi u(s \vec x) \cdot
  \vec x \; .
\end{equation}
As a result, the mass \textit{inside a given volume\/} $V$ will
change by
\begin{equation}
  \label{eq:11}
  \delta M(V) = \int_V \delta\rho(\vec x) \, \diff V \; .
\end{equation}
From Eq.~\eqref{eq:10} we obtain
\begin{align}
  \label{eq:12}
  \delta M(V) = {} & \delta \rho_0 \int_V u(s \vec x) \, \diff V
  \notag\\ 
  & {} + \delta \rho_0 \frac{\diff s}{\diff \rho_0} \rho_0 \int_V
  \nabla_\xi u(s \vec x) \cdot \vec x \, \diff V \; .
\end{align}
The first term on the r.h.s.\ of the previous equation is simply
$\delta \rho_0 M(V) / \rho_0$.  For the second term, we can apply the
general relation $\nabla(f \vec v) = \nabla f \cdot \vec v + f \nabla
\cdot \vec v$.  Thus we obtain
\begin{align}
  \label{eq:13}
  \delta M(V) & {} =  \frac{\delta \rho_0 M(V)}{\rho_0} + \delta
    \rho_0 \frac{\diff s}{\diff \rho_0} \frac{\rho_0}{s} \int_V \bigl[
    \nabla \cdot \bigl( u(s \vec x) \vec x \bigr) \notag\\
    &\phantom{{} =} {} - u(s \vec x) \nabla \cdot \vec x \bigr] \,
    \diff V \notag\\ 
    &{} = \frac{\delta\rho_0 M(V)}{\rho_0} + \delta \rho_0 \frac{\diff
    \ln s}{\diff \rho_0} \biggl[ \oint_{\partial V} \rho_0 u(s \vec x)
    \vec x \cdot \vec n \, \diff S \notag\\
    &\phantom{{} =} {} - 3 \int_V \rho_0 u(s \vec x) \, \diff V
    \biggr] \; ,
\end{align}
where we have used the Gauss theorem ($\partial V$ is the boundary of
$V$, $\vec n$ is the unit vector associated with the oriented surface
element).  Since $\rho$ is constant on $\partial V$, we find
\begin{align}
  \label{eq:14}
  \delta M(V) &{} = \frac{\delta \rho_0 M(V)}{\rho_0} + \frac{\delta
    \rho_0}{2 \rho_0} \biggl[ \rho_\mathrm{b} \int_V \nabla \cdot \vec
  x \, \diff V - 3 M(V) \biggr] \notag\\
  &{} = -\frac{\delta \rho_0 M(V)}{2 \rho_0} + \frac{3 \delta \rho_0
    \rho_\mathrm{b} V}{2 \rho_0} \; .
\end{align}
Therefore, for perturbations at constant cloud mass a variation $\delta
\rho_0$ of the ``central'' density must be compensated for by a
suitable change of volume $\delta V$.  Since the change of mass due to
a change of volume $\delta V$ is $\rho_\mathrm{b} \delta V$, we must
have
\begin{equation}
  \label{eq:15}
  -\frac{\delta \rho_0 M}{2 \rho_0} + \frac{3 \delta \rho_0
  \rho_\mathrm{b} V}{2 \rho_0} + \delta V \rho_\mathrm{b} = 0 \; .
\end{equation}
This equation controls the change of central density $\delta \rho_0$
when the volume of the cloud changes.

\subsection{Total mass in virial equilibrium}
\label{sec:total-mass}

For the following calculations it is useful to obtain a simple
alternative expression for the total mass applicable to a
self-gravitating isothermal cloud in hydrostatic equilibrium.  For the
purpose, we will use Eq.~\eqref{eq:6} in the integral of the mass:
\begin{align}
  \label{eq:16}
  M &{} = \int_V \rho(\vec x) \, \diff V = -\frac{c_\mathrm{s}^2}{4
  \pi G} \int_V \nabla \cdot \left( \frac{\nabla \rho}{\rho} \right)
  \, \diff V \notag\\
  &{} = - \frac{c_\mathrm{s}^2}{4 \pi G} \oint_{\partial V} \frac{\nabla
  \rho \cdot \vec n}{\rho} \, \diff S \; .
\end{align}
This expression can be simplified by noting that $\rho$ is constant on
$\partial V$ and $\nabla \rho \cdot \vec n = -\| \nabla \rho \|$,
because the boundary is defined as an iso-density surface.  [Here $\|
\vec A \| = \sqrt{\vec A \cdot \vec A}$ denotes the norm of the vector
$\vec A$.]\@ Hence, we find
\begin{equation}
  \label{eq:17}
  M = \frac{c_\mathrm{s}^2 \rho_0 s}{4 \pi G \rho_\mathrm{b}}
  \oint_{\partial V} \bigl\| \nabla_\xi u \bigr\| \, \diff S \; .
\end{equation}
The use of this relation is equivalent to the use of the equation for
virial equilibrium.

\subsection{Compressibility}
\label{sec:compressibility}

We now evaluate the change of the external pressure of a cloud as the
cloud volume changes, i.e.\ the derivative $\partial p_\mathrm{b} /
\partial V$ at constant temperature, constant mass, and fixed shape.

If we allow for changes of size, at the boundary Eq.~\eqref{eq:10}
becomes
\begin{align}
  \label{eq:18}
  \delta \rho_\mathrm{b} = {}& \frac{\delta \rho_0
    \rho_\mathrm{b}}{\rho_0} + \rho_0 \frac{\diff s}{\diff \rho_0}
  \delta \rho_0 \nabla_\xi u(s \vec x) \cdot \vec x \notag\\
  & {} - \frac{\rho_0 s \delta V}{\oint_{\partial V} \bigl\| \nabla_\xi
    u(s \vec x) \bigr\|^{-1} \, \diff S} \; .
\end{align}
The last term on the right-hand-side of this equation can be explained
as follows.  Since the cloud can change its size, we expect a related
variation of the external density $\rho_\mathrm{b}$.  Such change can
be written as $\delta \tilde\rho = -\bigl\| \nabla \rho(\vec x)
\bigr\| \delta x$ for each point $\vec x$, where $\delta x$ is the
``stretch'' of the boundary at $\vec x$ (we recall that this stretch
is always parallel to $\nabla \rho$).  We also note that $\delta
\tilde\rho$ is the same for all points on the boundary of the cloud,
because by definition this quantity can be identified with $\delta
\rho_\mathrm{b}$ in the case where $\delta \rho_0 = 0$.  Thus the
volume changes by
\begin{equation}
  \label{eq:19}
  \delta V = \oint_{\partial V} \delta x \, \diff S = -\delta
  \tilde\rho \oint_{\partial V} \frac{\diff S}{\| \nabla \rho \|} \; .
\end{equation}
Inverting this equation, we obtain that the change of density due to a
modification of the size of the cloud is given by
\begin{equation}
  \label{eq:20}
  \delta \tilde\rho = -\frac{\delta V}{\oint_{\partial V} \| \nabla \rho
  \|^{-1} \, \diff S} = -\frac{\rho_0 s \delta V}{\oint_{\partial V}
  \bigl\| \nabla_\xi u(s \vec x) \bigr\|^{-1} \, \diff S} \; ,
\end{equation}
which is the last term of Eq.~\eqref{eq:18}.

The second term on the r.h.s.\ of Eq.~\eqref{eq:18} is constant for
any point $\vec x$ on the boundary $\partial V$ because all the other
terms in the equation are so.  Thus we can average this constant term
on the boundary using a convenient weighted mean:
\begin{align}
  \label{eq:21}
  \nabla_\xi u(s \vec x) \cdot \vec x &{} = \oint_{\partial V}
  \!\frac{\nabla_\xi u(s \vec x) \cdot \vec x}{\bigl\| \nabla_\xi u(s
    \vec x) \bigr\|} \, \diff S \!\biggm/\! \oint_{\partial V}
  \!\frac{\diff S}{\bigl\| \nabla_\xi u(s \vec x) \bigr\|} \notag\\ 
  &{} = -\oint_{\partial V} \vec x \cdot \vec n \, \diff S \!\biggm/\!
  \oint_{\partial V} \frac{\diff S}{\bigl\| \nabla_\xi u(s \vec x)
    \bigr\|} \notag\\
  &{} = -\frac{3 V}{\oint_{\partial V} \bigl\| \nabla_\xi u(s \vec x)
    \bigr\|^{-1} \, \diff S} \; .
\end{align}
With this device the second term is reduced to a form that is similar
to that of the last term of Eq.~\eqref{eq:18}.

\subsection{Generalized Boyle's law}
\label{sec:gener-boyl-law}

Let us briefly summarize the main results obtained so far.  We 
wish to obtain an expression for $\partial \rho_\mathrm{b} / \partial
V$ at constant temperature and mass or, equivalently, an expression
for $\delta \rho_\mathrm{b}$ as a function of $\delta V$.
Equation~\eqref{eq:18} is close to our need, but, unfortunately,
it contains the auxiliary quantity $\delta\rho_0$.  On the other hand, we
can use Eq.~\eqref{eq:15} to express $\delta\rho_0$ in terms of
$\delta V$ and thus eliminate $\delta\rho_0$ from Eq.~\eqref{eq:18}.
Following this procedure, we obtain the desired $\partial
\rho_\mathrm{b} / \partial V$, but the resulting expression contains
integrals that are, at first sight, difficult to interpret.  

We now note that, with the help of Eq.~\eqref{eq:17}, a key term
involving an integration of $\bigl\| \nabla_\xi u \bigr\|$ (see
Eqs.~\eqref{eq:18} and \eqref{eq:21}) can actually be recognized to be
proportional to $GM/V^{4/3}$:
\begin{equation}
  \label{eq:22}
  \frac{1}{\oint_{\partial V} \bigl\| \nabla_\xi u(s \vec x)
  \bigr\|^{-1} \, \diff S} = \left( \frac{4\pi}{3} \right)^{1/3}
  \frac{M G \mu \rho_\mathrm{b}}{3 c_\mathrm{s}^2 \rho_0 s V^{4/3}} \; .
\end{equation}
This expression is reminiscent of what enters in the discussion of the
spherical cloud by Bonnor (1956).  For the purpose, we have introduced
the dimensionless \textit{form factor\/} as follows:
\begin{align}
  \label{eq:23}
  \mu = {}& 12 \pi \left( \frac{3}{4 \pi} \right)^{1/3}
  \frac{V^{4/3}}{S^2} \biggl( \frac{1}{S} \oint_{\partial V} \bigl\|
  \nabla_\xi u(s \vec x) \bigr\| \, \diff S
  \biggr)^{-1} \notag\\
  & {} \times \biggl( \frac{1}{S} \oint_{\partial V} \bigl\| \nabla_\xi u(s
  \vec x) \bigr\|^{-1} \, \diff S \biggr)^{-1} \; .
\end{align}
In this way, we have isolated the contribution from integrals of
$\bigl\| \nabla_\xi u \bigr\|$.  We postpone a description of the form
factor to the next subsection, where the reason for the numerical
factors adopted in its definition will become apparent.

Using relation \eqref{eq:22} in Eq.~\eqref{eq:21} and combining it
with Eqs.~\eqref{eq:18} and \eqref{eq:15} we find
\begin{equation}
  \label{eq:24}
  \delta \rho_\mathrm{b} = -\frac{2 \rho_\mathrm{b}^2 \delta V}{3
    \rho_\mathrm{b} V - M} \left[ 1 - \left( \frac{4\pi}{3}
    \right)^{1/3} \frac{G \mu M^2}{6 c_\mathrm{s}^2 \rho_\mathrm{b}
    V^{4/3}} \right] \; . 
\end{equation}
Finally, using Eq.~\eqref{eq:5} we obtain the desired result:
\begin{equation}
  \label{eq:25}
  \left( \frac{\partial p_\mathrm{b}}{\partial V} \right)_{T, M} =
  -\frac{2 p_\mathrm{b}}{3 V} \frac{\displaystyle{1 - \left(
        \frac{4\pi}{3} \right)^{1/3} \frac{G \mu M^2}{6 p_\mathrm{b}
        V^{4/3}}}}{\displaystyle{1 - \frac{M c_\mathrm{s}^2}{3
        p_\mathrm{b} V}}} \; . 
\end{equation}
This expression generalizes Eq.~(2.16) in the article by Bonnor
(1956).  Before discussing the physical interpretation of the equation
just found and its consequences, the form factor $\mu$ deserves a
special digression.

\subsection{Form factor}
\label{sec:form-factor}

The form factor, which plays a key role in the equilibrium and
stability of the cloud, is defined in terms of integrals of $\bigl\|
\nabla_\xi u \bigr\|$ and $\bigl\| \nabla_\xi u \bigr\|^{-1}$ on the
boundary $\partial V$.  Actually, the complex definition \eqref{eq:23}
has a simple interpretation in terms of general properties of $\mu$.

\subsubsection{$\mu$ is dimensionless}
\label{sec:mu-dimensionless}

As already noted, $\mu$ is a dimensionless quantity, which is a choice
that allows us to preserve the scaling of the terms as found in
Bonnor's (1956) derivation.  In this respect, note that the term
$V^{4/3}$ compensates for the two surface integrals in
Eq.~\eqref{eq:23}.  As shown below, the introduction of a
dimensionless quantity allows us to characterize and summarize the
properties associated only with the shape of the cloud.

\subsubsection{$\mu$ is scale invariant}
\label{sec:mu-scale-invariant}

It is easily verified that the form factor $\mu$ is scale invariant,
i.e.\ it does not change under the transformation $\vec x \mapsto t
\vec x$, with $t > 0$.  In other words, the form factor depends on the
\textit{shape\/} of the cloud but does not depend on its
\textit{size}.  Actually, the scale invariance is directly related to
the fact that $\mu$ is dimensionless and to the vectorial notation
used in its definition.

\subsubsection{$\mu = 1$ for a sphere}
\label{sec:mu-=-1}

In the case of a sphere, the form factor reduces to unity.  In fact,
in this case by symmetry $\bigl\| \nabla_\xi u(s \vec x) \bigr\|$ is
constant on the surface of the sphere, so that
\begin{equation}
  \label{eq:26}
  \mu = 12 \pi \left( \frac{3}{4 \pi} \right)^{1/3}
  \frac{V^{4/3}}{S^2} = 1\; .
\end{equation}
Here $S$ is the area of the surface of the sphere.  Thus, from
Eq.~\eqref{eq:25}, we recover Bonnor's (1956) result for the sphere.
Note that the numerical factors in the definition of
$\mu$  have been chosen precisely to recover $\mu = 1$ for the sphere.

\subsubsection{$\mu$ never exceeds unity}
\label{sec:mu-le-1}

The form factor can be thought of as the product of two terms, $\mu =
\mu_1 \mu_2$, with (cf.\ Eq.~\eqref{eq:26})
\begin{equation}
  \label{eq:27}
  \mu_1 = 12 \pi \left( \frac{3}{4 \pi} \right)^{1/3}
  \frac{V^{4/3}}{S^2} \; .
\end{equation}
The first term, $\mu_1$, compares the volume $V$ of the cloud with its
surface $S$.  This quantity equals unity when the cloud is a sphere
and is always smaller than unity in the non-spherical case.  In fact,
the sphere is the solid which has the largest volume at given surface
(this intuitive theorem has a far from trivial proof, due to De~Giorgi
1958).  The second term, $\mu_2$, is the ratio between the generalized
harmonic mean of $\bigl\| \nabla_\xi u \bigr\|$ on the boundary
$\partial V$ and the generalized simple mean of the same quantity.
The harmonic mean is always smaller than the simple mean, the two
being equal only when all the values involved are constant.  Thus
$\mu_2 \le 1$ and $\mu_2 = 1$ if and only if $\bigl\| \nabla_\xi u
\bigr\|$ is constant on $\partial V$ (this happens for the isothermal
sphere).  In conclusion, $\mu \le 1$ in general and $\mu = 1$ only for
the sphere.

\section{The Bonnor instability}
\label{sec:bonnor-instability}

Equation \eqref{eq:25} allows us to discuss the Bonnor instability
(often known as gravothermal catastrophe; see Lynden-Bell \& Wood
1968) for clouds of arbitrary shape.  This instability, we recall, is
due to a change in the sign of the derivative of $p_\mathrm{b}$ with
respect to $V$.  If this derivative is negative, as normally happens
when gravity can be neglected, a slight decrease of the volume will
produce an increase of the cloud boundary pressure, which would tend
to restore the initial configuration.  If, instead, this derivative is
positive, then a small decrease in the volume of the cloud would
correspond to a reduction of the internal boundary pressure and thus
to a collapse of the cloud.

\begin{figure}[!t]
  \centerline{\resizebox{\hsize}{!}{\includegraphics{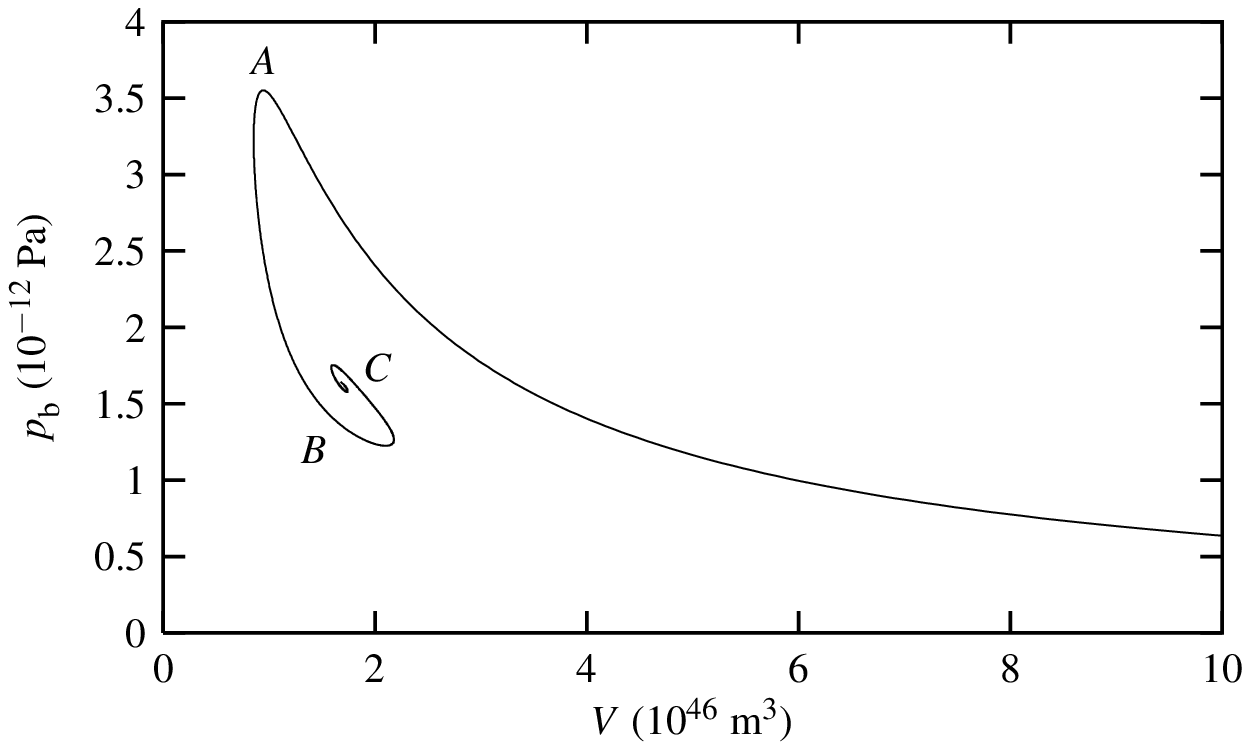}}}
  \caption{The $p$-$V$ plot for an isothermal spherical cloud made of
    molecular hydrogen at $T = 10 \mbox{ K}$ ($c_\mathrm{s} \simeq 200
    \mbox{ m s}^{-1}$) with mass $M = 1 \mbox{ M}_\odot$.  The
    isothermal lies on a simple $pV = \mbox{const}$ curve at large
    volumes, but then at smaller sizes departs from it and the cloud
    becomes unstable.  The region of stability is on the right side of
    point $A$ (which corresponds to the maximum of the boundary
    pressure $p_\mathrm{b}$).  The solution spirals into the critical
    point $C$.
    \label{fig:1}}
\end{figure}

In order to illustrate the instability process in more detail, let us
first refer to the spherical case.  We have integrated the
differential equation \eqref{eq:25} numerically under the boundary
condition $p_\mathrm{b} V = c_\mathrm{s}^2 M$ for $V \rightarrow
\infty$.  Figure~\ref{fig:1} shows the case of a spherical cloud of
molecular hydrogen at $T = 10 \mbox{ K}$ ($c_\mathrm{s} \simeq 200
\mbox{ m s}^{-1}$) with mass $M = 1 \mbox{ M}_\odot$.  The plot
clearly shows that the cloud follows approximately the standard law
$pV = \mbox{const}$ at large volumes, but when sufficiently compressed
it exhibits the Bonnor instability.  In particular, the $p$-$V$ curve
has a maximum and then spirals into a singular point.  All points of
the $p$-$V$ curve where the derivative is positive correspond to
unstable equilibria, since a small decrease of the volume will make
the cloud collapse.  Bonnor has shown that actually, in the case of a
spherical cloud, all points in the $p$-$V$ plot between $A$ and the
critical point correspond to unstable equilibria (below we will
generalize this result to all clouds).  The idea is that a spherical
cloud with pressure and volume beyond $A$ will present an instability
for compressions of some internal ``core'' (we will discuss this
instability in the general case below).

\begin{figure}[!t]
  \centerline{\resizebox{\hsize}{!}{\includegraphics{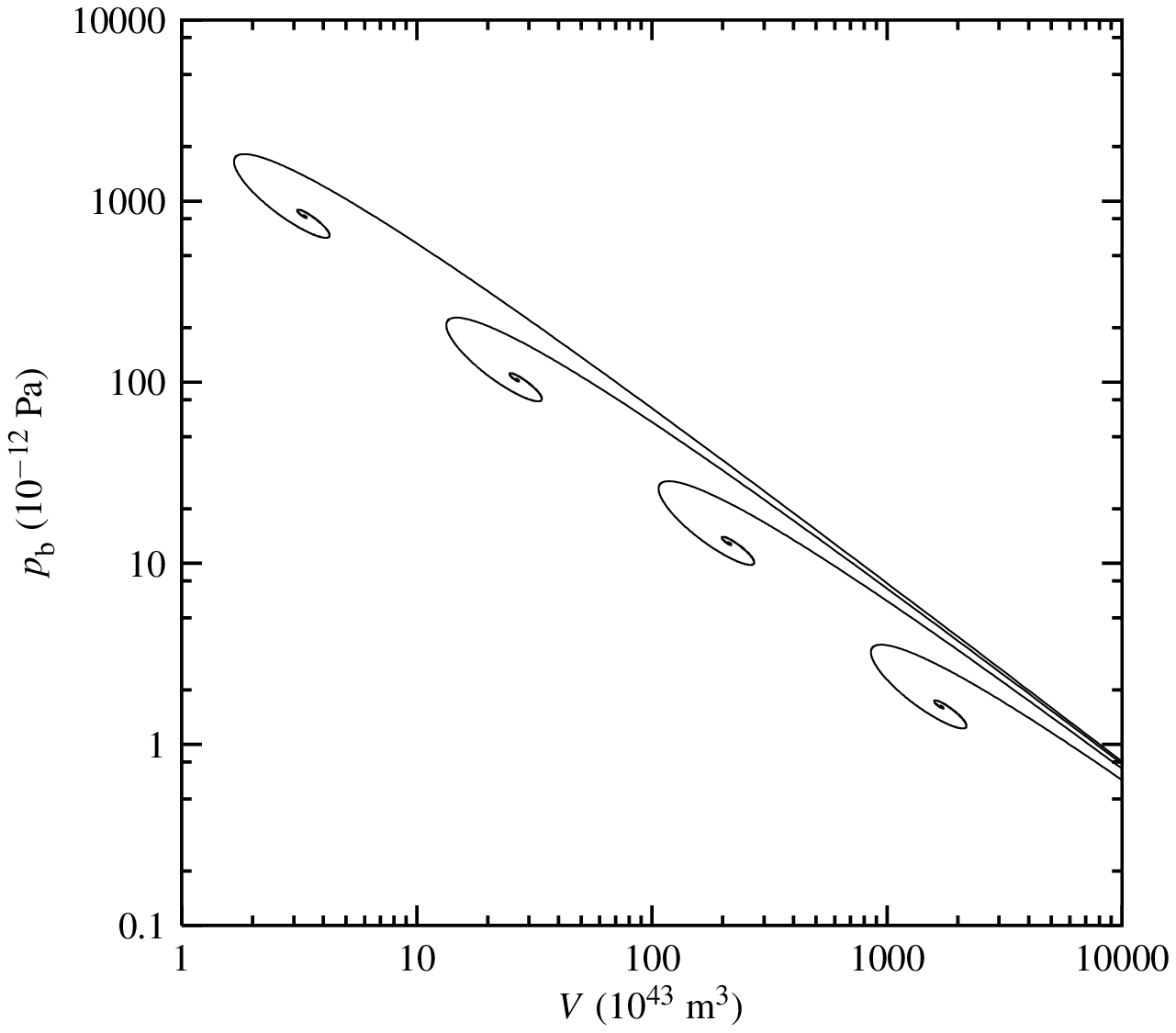}}}
  \caption{The $p$-$V$ plot (in logarithmic coordinates) for
    isothermal clouds of different shapes made of molecular hydrogen
    at $T = 10 \mbox{ K}$ with mass $M = 1 \mbox{ M}_\odot$.  The form
    factors $\mu$ are, from top left to bottom right, $0.125$, $0.25$,
    $0.5$, and $1$.  In logarithmic coordinates, a change of $\mu$ is
    described by a simple translation of the isothermal curve.
    \label{fig:2}}
\end{figure}

In the general case of an arbitrary shape, the situation turns out to
be similar.  In fact, going back to Eq.~\eqref{eq:25}, we can see that
the form factor enters the global equation of state together with the
gravitational constant $G$.  In other words, a non-spherical cloud
behaves like a spherical cloud with a \textit{reduced\/} gravitational
constant $G \mu$.  As discussed above (see
Sect.~\ref{sec:form-factor}), the isothermal sphere has the largest
admissible value for the form factor, $\mu = 1$, and thus for the
spherical shape the effects introduced by self-gravity are largest.
The $G \mu$ invariance also implies that the isothermal $p$-$V$ curves
exhibit similar shapes regardless of the value of $\mu$.  In
Fig.~\ref{fig:2} we have plotted the integral curves of
Eq.~\eqref{eq:25} under the assumption that the cloud form factor is
constant as $V$ changes.  Figure~\ref{fig:2} clearly shows that the
Bonnor instability occurs at any $\mu > 0$; moreover, the effect of
reducing the form factor is to \textit{postpone\/} the onset of the
instability to smaller volumes.  In the limit of a vanishing form
factor we recover the standard law $p V = \mbox{ const}$.  From this
figure we also see that in logarithmic coordinates a change of the
form factor $\mu$ simply results in a simple translation of the
$p$-$V$ curve.  In fact, it is easily checked that, if we define
\begin{align}
  \label{eq:28}
  \tilde p &{} = p \mu^3 \; , &
  \tilde V &{} = V \mu^{-3} \; ,
\end{align}
then Eq.~\eqref{eq:25} is transformed back into the equation
appropriate for a sphere, since $\mu$ drops out.  As a result, a curve
relative to a form factor $\mu$ can be obtained from the curve for the
sphere by multiplying the volume by $\mu^3$ and the pressure by
$\mu^{-3}$.

Point $A$ is a zero of $\partial p_\mathrm{b} / \partial V$ and thus
satisfies
\begin{equation}
  \label{eq:29}
  p_A = \left( \frac{4 \pi}{3} \right)^{1/3} \frac{G \mu M^2}{6
  V_A^{4/3}} \; .
\end{equation}
Using the Emden (1907) solution for the sphere, and taking advantage
of the $\mu^3$ scaling, we can give the position of point $A$ for
arbitrary shapes:
\begin{align}
  \label{eq:30}
  V_A & \simeq 0.290307 \left( \frac{G \mu M}{c_\mathrm{s}^2}
  \right)^3 \; , \\
  \label{eq:31}
  p_A & \simeq 1.39766 \frac{c_\mathrm{s}^8}{G^3 \mu^3 M^2} \; .
\end{align}
The curves in the $p$-$V$ plane spiral into the critical point $C$.
This point is easily obtained by imposing that both the numerator and
the denominator of Eq.~\eqref{eq:25} vanish.  Then we obtain
\begin{align}
  \label{eq:32}
  V_C & {} = \frac{\pi}{6} \left( \frac{G \mu M}{c_\mathrm{s}^2}
  \right)^3 \; , \\
  \label{eq:33}
  p_C & {} = \frac{2}{\pi} \frac{c_s^8}{G^3 \mu^3 M^2} \; .
\end{align}

Let us now complete the discussion of the Bonnor instability for a
general cloud.  We have already noted that all clouds for which the
derivative $\partial p_\mathrm{b} / \partial V$ is negative are
unstable.  Actually, we can show that all points in the $p$-$V$ curve
between $A$ and $C$ are unstable.  In fact, suppose that the cloud is
in a state represented by a point $B$ between $A$ and $C$, and let us
consider smaller and smaller portions of the cloud with boundaries
characterized by a constant pressure.  Let us call $V(p)$ and $M(p)$,
respectively, the volume and the mass of the part of the cloud with
pressure larger than $p$.  We now observe that a small compression of
this part of the cloud is still described by Eq.~\eqref{eq:25} with
the obvious replacements $p_\mathrm{b} \mapsto p$, $V \mapsto V(p)$,
and $M \mapsto M(p)$.  As we increase the value of $p$, we approach
the central part of the cloud.  Assuming that the maximum cloud
density $\rho_0$ is finite, we have $M(V) \le \rho_0 V$, and the
numerator of the last term of Eq.~\eqref{eq:25} will tend to unity
because $M^2 / V^{4/3} \le \rho_0^2 V^{2/3} \rightarrow 0$ as $V
\rightarrow 0$.  As a result, in the central part of the cloud the gas
follows the standard law $pV = \mbox{const}$, i.e.\ we are in the
right part of the plot $p$-$V$.  If we now move back from the central
part to the boundary of the cloud represented by point $B$ we must
travel along the $p$-$V$ curve (actually, we are moving along
different $p$-$V$ curves, with respect to the reference one, because
the mass $M(V)$ of the sub-cloud changes while we consider larger and
larger volumes) and we will reach, at a certain boundary pressure
$p_\mathrm{a}$, a point of type $A$.  In other words, we have shown
that the sub-cloud characterized by $p > p_\mathrm{a}$ is on a point
of type $A$ of the $p$-$V$ plot.  If this sub-cloud has a small
negative fluctuation of its volume $V_\mathrm{a} = V(p_\mathrm{a})$, a
decrease of its boundary pressure will occur and this will lead to a
further reduction of its volume.  Thus we must conclude that every
cloud characterized by a point in the $p$-$V$ plot between $A$ and $C$
is unstable.  This is the generalization of the Bonnor instability to
non-spherical clouds.

Point $A$ then defines the boundary of the stability region.  If the
external pressure is larger than $p_A$, no equilibrium solution exists
for the cloud.  This provides an upper limit for the cloud mass:
\begin{equation}
  \label{eq:34}
  M < M_\mathrm{crit} \simeq 1.18223 \frac{c_\mathrm{s}^4}{\sqrt{G^3
  \mu^3 p_\mathrm{ext}}} \; ,
\end{equation}
where $p_\mathrm{ext}$ is the external pressure (that we suppose to be
balanced by the cloud boundary pressure $p_\mathrm{b}$).  Inserting
here a typical value for the pressure of the interstellar medium,
$p_\mathrm{ext} = 1.4 \times 10^{-13} \mbox{ Pa}$, and using
$c_\mathrm{s} = 200 \mbox{ m s}^{-1}$ we find $M_\mathrm{crit} \simeq
4.7 \mu^{-3/2} \mbox{ M}_\odot$.

If condition \eqref{eq:34} is met, we can still be in the instability
region if we are in the spiral part of the $p$-$V$ curve.  This
happens if $p_\mathrm{b} V < p_A V_A$, and thus a necessary condition
for stability, independent of $\mu$, is
\begin{equation}
  \label{eq:35}
  \bar \rho < \rho_\mathrm{crit} \simeq 2.46457
  \frac{p_\mathrm{ext}}{c_\mathrm{s}^2} \; , 
\end{equation}
where $\bar \rho = M/V$ is the mean density of the cloud.  Inserting
here the typical values used before, we find $\rho_\mathrm{crit}
\simeq 8.6 \times 10^{-18} \mbox{ kg m}^{-3}$, corresponding to about
$2600$ molecules per $\mbox{cm}^3$.  Equation~\eqref{eq:35} can also
be read in terms of a density contrast requirement $\bar \rho <
2.46457 \rho_\mathrm{b}$.  \textit{Equation \eqref{eq:35} proves that
  the condition for the Bonnor instability can be cast in a form that
  is independent of the shape of the cloud}.  However, Bonnor's
stability result for spheres, when expressed as a condition $\rho_0 /
\rho_\mathrm{b} < 14.0$, does not hold for clouds of arbitrary shape.

Finally, we wish to emphasize that Eqs.~\eqref{eq:34} and
\eqref{eq:35} are only \textit{necessary\/} conditions for equilibrium
and stability, and \textit{not sufficient\/} conditions.  In other
words, other kinds of instabilities may occur for clouds that satisfy
Eqs.~\eqref{eq:34} and \eqref{eq:35}.  Moreover, the presence of
rotation or magnetic fields may qualitatively change these results, as
suggested by recent investigations on the equilibrium of clouds in the
presence of magnetic fields (see Fiege \& Pudritz 2000; Curry \&
Stahler 2001).

\section{Boyle's law for (infinite) cylinders}
\label{sec:boyles-law-cylinders}

So far we have assumed that the cloud has finite volume.  However,
several observed clouds appear as filaments and can be well modeled by
cylinders.  The technique described above, suitably adapted, can be
used also to obtain Boyle's law for (infinite) cylindrical clouds of
arbitrary shape, described by a mass density of the form
\begin{equation}
  \label{eq:36}
  \rho(x_1, x_2, x_3) = \rho(x_1, x_2) \; .
\end{equation}
Note that no assumption on the function $\rho(x_1, x_2)$ has been
made, and in particular we \textit{do not\/} assume a circular
section.  We can now calculate the compressibility $\partial
p_\mathrm{b} / \partial V$ following the steps taken earlier in this
paper.  Since the calculations are very similar to the 3D case, we
will just outline the main differences here.

Equations \eqref{eq:3} to \eqref{eq:10} remain unchanged; however,
since all functions are independent of $x_3$, it is convenient to
define here $\vec x = (x_1, x_2)$, $\vec\xi = (\xi_1, \xi_2)$, and
$\nabla = (\partial / \partial x_1, \partial / \partial x_2)$.  The
``total'' mass cannot be defined for infinite cylinders, and hence we
will refer to a linear mass density $\lambda$ defined as
\begin{equation}
  \label{eq:37}
  \lambda = \int_S \rho(\vec x) \, \diff S \; ,
\end{equation}
where $S$ is the section of the cylinder.

\subsection{Perturbations at constant mass}
\label{sec:constance-mass-1}

Using Eq.~\eqref{eq:10}, which remains unchanged in the 2D case, we can
write the change of ``mass'' $\delta\lambda(S)$ when the central
density changes from $\rho_0$ to $\rho_0 + \delta\rho_0$ assuming no
variation in the cloud section $S$.  The result obtained is basically
the same as in Sect.~\ref{sec:constance-mass} with the important
difference that now $\nabla \cdot \vec x = 2$.  As a result,
Eq.~\eqref{eq:14} now becomes
\begin{align}
  \label{eq:38}
  \delta \lambda(S) & {} = \frac{\delta \rho_0 \lambda(S)}{\rho_0} +
  \frac{\delta \rho_0}{2 \rho_0} \biggl[ \rho_\mathrm{b} \int_S
  \nabla \cdot \vec x \, \diff S - 2 \lambda(S) \biggr] \notag\\
  &{} = \frac{\delta \rho_0 \rho_\mathrm{b} S}{\rho_0} \; .
\end{align}
Note that the two terms involving $\lambda(S)$ cancel each other.  If
we now allow for a change of section $\delta S$ and impose that the
total ``mass'' $\lambda$ be constant we find the analogue of
Eq.~\eqref{eq:15}:
\begin{equation}
  \label{eq:39}
  \frac{\delta \rho_0}{\rho_0} + \frac{\delta S}{S} = 0 \; .
\end{equation}
This equation is independent of $\rho_\mathrm{b}$ and simply implies
$\rho_0 \propto 1/S$.

\subsection{Compressibility}
\label{sec:compressibility-1}

Equation~\eqref{eq:18} rewritten for cylinders is
\begin{align}
  \label{eq:40}
  \delta \rho_\mathrm{b} = {}& \frac{\delta \rho_0
    \rho_\mathrm{b}}{\rho_0} + \rho_0 \frac{\diff s}{\diff \rho_0}
  \delta \rho_0 \nabla_\xi u(s \vec x) \cdot \vec x \notag\\
  &{} - \frac{\rho_0 s \delta S}{\oint_{\partial S} \bigl\| \nabla_\xi
    u(s \vec x) \bigr\|^{-1} \, \diff \ell}
\end{align}
Similarly to the 3D case, we can rewrite the second term on the
r.h.s.\ in a more convenient way by using a suitable average.  In this
case we have
\begin{align}
  \label{eq:41}
  \nabla_\xi u(s \vec x) \cdot \vec x &{} = \oint_{\partial S}
  \!\frac{\nabla_\xi u(s \vec x) \cdot \vec x}{\bigl\| \nabla_\xi u(s
    \vec x) \bigr\|} \, \diff \ell \!\biggm/\! \oint_{\partial S}
  \!\frac{\diff \ell}{\bigl\| \nabla_\xi u(s \vec x) \bigr\|} \notag\\
  &{} = -\frac{2 S}{\oint_{\partial S} \bigl\| \nabla_\xi u(s \vec x)
    \bigr\|^{-1} \, \diff \ell} \; .
\end{align}
Inserting this result in Eq.~\eqref{eq:40} and using relation
\eqref{eq:39} we find
\begin{equation}
  \label{eq:42}
  \delta\rho_\mathrm{b} = -\frac{\delta S \rho_\mathrm{b}}{S} \; .
\end{equation}
This simple result should be compared with Eq.~\eqref{eq:24}, valid in
the three-dimensional case.  We finally find the analogue of
Eq.~\eqref{eq:25}, 
\begin{equation}
  \label{eq:43}
  \left( \frac{\partial p_\mathrm{b}}{\partial S} \right)_{T, \lambda} =
  -\frac{p_\mathrm{b}}{S} \; .
\end{equation}
This equation \textit{states that a self-gravitating isothermal
  cylinder behaves like a normal gas under compressions}, i.e.\ it
follows the standard law $p_\mathrm{b} S = \mbox{ const}$.  However,
we should emphasize that in general the constant that appears on the
r.h.s.\ of this expression is not equal to $\lambda c_\mathrm{s}^2$.
In other words, an isothermal cylinder at temperature $T$ follows a
law similar to the one of a perfect gas but \textit{at an effective
  temperature\/} $T_\mathrm{eff} < T$ (see Eq.~\eqref{eq:2}; see also
Sect.~\ref{sec:circular-cylinder} below).  The difference $T -
T_\mathrm{eff}$ will depend on the particular shape of the cylinder
and on its linear mass density $\lambda$.  Furthermore, for cylinders,
since no form factor needs to be introduced, Eq.~\eqref{eq:43} can be
integrated exactly without additional assumptions.

Since the cylinder follows a law similar to the standard law for
perfect gases, no Bonnor instability is present in this case.  Still
we have to make sure that $T_\mathrm{eff} > 0$.  From
Eq.~\eqref{eq:43} we see that this is true as long as an equilibrium
exists with $p_\mathrm{b} > 0$.  In the following subsection we will
consider an axisymmetric cylinder in order to illustrate this point by
means of a simple example.

\subsection{Axisymmetric cylinder}
\label{sec:circular-cylinder}

The case of a circular cylinder, with density of the form $\rho(\vec
x) = \rho(r)$ dependent only on the distance $r$ from the axis, has
already been the subject of several investigations (see, e.g.,
Ostriker 1964; Horedt 1986; Bastien et al.\ 1991).  For an
axisymmetric cylinder the \textit{exact\/} form of the solution of
Emden's equation is known (Stodo\l kiewicz 1963; Ostriker 1964).  In our
notation, the solution can be written as
\begin{equation}
  \label{eq:44}
  \rho(r) = \rho_0 \left[ 1 + \frac{(s r)^2}{8} \right]^{-2} \; .
\end{equation}
We can integrate this expression out to a certain radius $R$ in order
to get the linear density $\lambda$ of the cylinder.  On the other
hand, if we know the linear density $\lambda$ and the radius $R$ of the
cloud, we can obtain the central density $\rho_0$.  A simple
calculation yields
\begin{equation}
  \label{eq:45}
  \rho_0 = \frac{\lambda}{\displaystyle{S - \frac{G \lambda S}{2
  c_\mathrm{s}^2}}} \; .
\end{equation}
As anticipated in Eq.~\eqref{eq:39}, we find $\rho_0 \propto 1/S$.
Inserting this in Eq.~\eqref{eq:44} at $r = R$ and using
Eq.~\eqref{eq:5} we finally find
\begin{equation}
  \label{eq:46}
  p_\mathrm{b} S = \lambda c_\mathrm{s}^2 \left( 1 - \frac{G \lambda}{2
  c_\mathrm{s}^2} \right) \; .
\end{equation}
This is a standard isothermal law, equivalent to Boyle's law, but
characterized by an ``effective'' temperature
\begin{equation}
  \label{eq:47}
  T_\mathrm{eff} = T - \frac{G \lambda m}{2 k} \; .
\end{equation}
The term $G \lambda m / 2 k$ represents the contribution from the
potential energy of the cylinder, as indicated by Eq.~\eqref{eq:2}.
From Eq.~\eqref{eq:46} we immediately see that when the cloud linear
density is larger than a critical density $\lambda_\mathrm{crit} = 2
c_s^2 / G$, the cloud cannot be in equilibrium (the required boundary
pressure $p_\mathrm{b}$ would be negative).

\section{Conclusions}
\label{sec:conclusions}

The main results obtained in this paper can be summarized as follows.
\begin{enumerate}
\item The generalized Boyle's law for an isothermal self-gravitating
  cloud in hydrostatic equilibrium has been studied.  No assumption
  has been made on the shape of the cloud.
\item An expression for the compressibility $\partial p_\mathrm{b} /
  \partial V$ has been obtained and, from that, two conditions for
  stable equilibrium have been derived.
\item The equilibrium of the cloud has been shown to be controlled by
  one dimensionless form factor.  The spherical shape turns out to be
  the one for which the effect of self-gravity is largest.
\item A simple relationship between the isothermal curves in the
  $p$-$V$ plane and the form factor has been obtained.
\item The actual condition for the Bonnor instability can be written
  in a form that does not depend on the shape of the cloud.
\item For infinite cylinders of arbitrary shape we have shown that the
  Bonnor instability is suppressed.
\end{enumerate}

\begin{acknowledgements}
  We would like to thank Jo\~ ao Alves for helpful discussions.  Part
  of this work has been carried out at the Scuola Normale Superiore,
  Pisa, Italy.
\end{acknowledgements}

\end{document}